\documentclass[twocolumn,showpacs,superscriptaddress,floatfix,aps,prl]{revtex4-1}
\usepackage[english]{babel}
\usepackage{amsmath}
\usepackage{graphicx}
\usepackage[colorinlistoftodos]{todonotes}
\usepackage{soul}
\usepackage{amsmath}
\usepackage{amssymb}
\usepackage{natbib}
\usepackage{color}
\usepackage[caption=false]{subfig}
\usepackage{floatrow}

\begin{document}

\author{N. Bachelard}
\affiliation{Institut Langevin, ESPCI ParisTech, CNRS, 1 rue Jussieu, 75238 Paris Cedex 05, France}

\author{C. Garay}
\affiliation{Institut Langevin, ESPCI ParisTech, CNRS, 1 rue Jussieu, 75238 Paris Cedex 05, France}

\author{J. Arlandis}
\affiliation{Institut Langevin, ESPCI ParisTech, CNRS, 1 rue Jussieu, 75238 Paris Cedex 05, France}

\author{R. Touzani}
\affiliation{Laboratoire de Math\'ematiques, Universit\'e Blaise Pascal, CNRS (UMR 6620), 63177 Aubi\`ere Cedex, France}

\author{P. Sebbah}
\email[Contact: ]{patrick.sebbah@espci.fr}
\affiliation{Institut Langevin, ESPCI ParisTech, CNRS, 1 rue Jussieu, 75238 Paris Cedex 05, France}

\date{\today}

\title{Coalescence of Anderson-localized modes at an exceptional point in 2D random media}

\begin{abstract}
In non-hermitian systems, the particular position at which two eigenstates coalesce under a variation of a parameter in the complex plane is called an exceptional point.
A non-perturbative theory is proposed which describes the evolution of modes in 2D open dielectric systems when permittivity distribution is modified. 
We successfully test this theory in a 2D disordered system to predict the position in the parameter space of the exceptional point between two Anderson-localized states.
We observe that the accuracy of the prediction depends on the number of localized states accounted for.
Such an exceptional point is experimentally accessible in practically relevant disordered photonic systems.

\end{abstract}

\pacs{42.25.Bs, 71.23.An, 71.70.Gm}

\maketitle

Losses are inherent to most physical systems, either because of dissipation or as a result of openness. These systems are described mathematically by a non-hermitian Hamiltonian, where eigenvalues are complex and eigenstates form a nonorthogonal set. In such systems, interaction between pairs of eigenstates when a set of external parameters is varied is essentially driven by the existence of exceptional points (EP). 
At an EP, eigenstates coalesce: Complex eigenvalues degenerate and spatial distributions become collinear. 
In its vicinity, eigenvalues display a singular topology \cite{Heiss1990} and encircling the EP in the parameter space leads to a residual geometrical phase \cite{Whitney2003,Carollo2003}.
Since their introduction by Kato in 1966 \cite{Kato}, EPs have turned to be involved in a rich variety of physical effects: Level repulsion \cite{Heiss1999}, mode hybridization \cite{Kammerer2008}, quantum phase transition \cite{Heiss2002}, lasing mode switching \cite{Liertzer2012a}, \textit{PT} symmetry breaking \cite{Klaiman2008,Guo2009a} or even strong coupling \cite{Choi2010}.
They have been observed experimentally in different systems such as microwave billiards \cite{Dietz2011}, chaotic optical microcavities \cite{Lee2009a} or two level atoms in high-Q cavities \cite{Choi2010}.
\\
Open random media are a particular class of non-hermitian systems. 
Here, modal confinement may be solely driven by the degree of scattering. 
For sufficiently strong scattering, the spatial extension of the modes becomes smaller than the system size, resulting in transport inhibition and Anderson localization \cite{Anderson1958}.
Disordered-induced localized states have raised increasing interest. They provide with natural optical cavities in random lasers \cite{Milner2005,Yang2011}.
They recently appeared to be good candidate for cavity QED \cite{Sapienza2010,Gao2013}, with the main advantage of being inherently disorder-robust.
These modes can be manipulated by a local change of the disorder and can be coupled to form necklace states \cite{Labonte2012,Bertolotti2005,Sebbah2006}, which open channels in a nominally localized system \cite{Pendry1987,Pendry1994}. These necklace states are foreseen as a key mechanism in the transition from localization to diffusive regime \cite{Vanneste2009}.
\textit{PT} symmetry has been studied in the context of disordered media and Anderson localization \cite{Bendix2009,Vemuri2011,Vazquez-Candanedo2014} but so far EPs between localized modes have not been investigated.

In this letter, coalescence at an EP between two Anderson-localized optical modes is demonstrated in a two dimensional (2D) dielectric random system.
To bring the system in the vicinity of an EP, the dielectric permittivity is varied at two different locations in the random system.
We first propose a general theory to follow the spectral and spatial evolution of modes in 2D dielectric open media. 
This theory is applied to the specific case of Anderson-localized modes to identify the position of an EP in the parameter space. 
This prediction is confirmed by Finite Element Method (FEM) simulations.
We show that this is a highly complex problem of multiple mode interaction where a large number of modes are involved.
We believe that our theory opens the way to a controlled local manipulation of the permittivity and the possibility to engineer the modes.
Furthermore, we think this approach can be easily extended to others kinds of networks e.g. coupled arrays of cavities \cite{Hartmann2006,Greentree2006}.

We first consider the general case of a finite-size dielectric medium in 2D space, with inhomogeneous dielectric constant distribution, $\epsilon(r)$.
In the frequency domain, the electromagnetic field follows the Helmholtz equation:
\begin{equation}
   \Delta E (r,\omega) + \epsilon(r) \omega^2  E(r,\omega) = 0
   \label{eq:Helmholtz}
\end{equation}
where $E(r,\omega)$ stands for the electrical field and the speed of light, $c$ = 1.
Eigensolutions of eq.~\eqref{eq:Helmholtz}, define the modes or eigenstates of the problem:
\begin{equation}
   (\Omega_i,|\Psi_i\rangle)_{i \in \mathbb{N}} \quad | \quad \Delta |\Psi_i\rangle +  \epsilon(r) \Omega_i^2 |\Psi_i\rangle = 0
   \label{eq:EigenStates}
\end{equation}
Because of its openness, the system has inherent losses, thus is described by a non-hermitian Hamiltonian.
For non-hermitian systems, modes are \textit{a priori} non-orthogonal, complex and their completeness is not ensured.
Here, we consider open systems with finite range permittivity $\epsilon(r)$ and where a discontinuity in the permittivity  provides a natural demarcation of the problem.
For problems fulfilling these two conditions, Leung \textit{et al.} \cite{P.T.LeungS.Y.LiuS.S.Tong1994,P.T.Leung1994,P.T.Leung1994,Leung1998} demonstrated the completeness of the set of eigenstates.
As a result, the electrical field can be expanded along the modes:
\begin{equation}
   E(r,\omega) = \sum_i a_i(\omega) |\Psi_i\rangle
   \label{eq:Expand}
\end{equation}
where $a_i(\omega)$ stand for expansion coefficients along the basis.
Moreover, if the eigenstates are not degenerated a biorthogonal product between modes can be defined \cite{Morse1953,Moiseyev1978}:
\begin{equation}
   \langle \Psi_p^{*}|\epsilon(r)|\Psi_q \rangle  = \delta_{pq}
   \label{eq:Projection}
\end{equation}

Now, we consider two locations $R_1$ and $R_2$ where the permittivity is varied
\begin{equation}
   \tilde{\epsilon}(r)=\epsilon(r) + \Delta\epsilon_1(r) p_1(r) + \Delta\epsilon_2(r) p_2(r)
   \label{eq:n}
\end{equation}
where $\left\{ p_i(r \in R_i) = 1 | p_i(r \notin R_i) = 0 \right\}_{i\in[1,2]}$ is the location and  $\left\{\Delta\epsilon_i(r)\right\}_{i\in[1,2]}$ the shape of the variation of permittivity.
Eq.~\eqref{eq:Helmholtz} becomes: 
\begin{equation}
   [\Delta + \omega^2 ( \epsilon(r) + \Delta\epsilon_1(r) p_1(r) +  \Delta\epsilon_2(r) p_2(r))] E(r,\omega)  =  0
   \label{eq:HelmholtzPertubated}
\end{equation} 
%%%%%%%%%%%%%%%%%%%%%%%
The permittivity distribution $\tilde{\epsilon}(r)$, describes a new disordered system with new modes $(\tilde{\Omega}_i,|\tilde{\Psi}_i\rangle)_{i\in \mathbb{N}}$.
Nevertheless, we can still use the basis of the original random system, $(\Omega_i,|\Psi_i\rangle)_{i \in \mathbb{N}}$, to expand the electric field as follows:
\begin{equation}
   E(r,\omega) = \sum_i b_i(\omega) |\Psi_i\rangle
   \label{eq:ExpandPert}
\end{equation}
where $b_i(\omega)$ are the new expansion coefficients.
Inserting eq.~\eqref{eq:ExpandPert} into eq.~\eqref{eq:HelmholtzPertubated}: \small
\begin{equation}
     \sum_i b_i(\omega) \left[\Delta + \omega^2 \left(\epsilon(r) + \Delta\epsilon_1(r) p_1(r) + \Delta\epsilon_2(r) p_2(r)\right) \right] |\Psi_i\rangle = 0
   \label{eq:interm}
\end{equation} \normalsize
Projecting eq.~\eqref{eq:interm} on $\langle \Psi_j^{*}|$, using eq.~\eqref{eq:EigenStates} and the biorthogonal product \eqref{eq:Projection} leads to:
\begin{equation}
   \forall\text{ \textit{i}} \quad b_i(\omega) \left( \Omega_i^{2}  - \omega^2 \right) = \omega^2 \sum_j b_j(\omega) C_{ij}
   \label{eq:HelmholtzPertubated_Proj}
\end{equation}
where
\begin{equation}
    C_{ij}  = \langle \Psi_j^*| \Delta \epsilon_1(r) p_1(r) |\Psi_i \rangle + \langle\Psi_j^*| \Delta \epsilon_2(r) p_2(r) |\Psi_i  \rangle
   \label{eq:Cij}
\end{equation}
If we consider a finite set of $N$ modes, the generalized eigenvalue problem of eq.~\eqref{eq:HelmholtzPertubated_Proj} can be written conveniently in a matrix form:
\begin{equation}
   \begin{pmatrix} \Omega_1^{2} & ... & 0 \\ \vdots & \ddots & \vdots \\ 0 & ... & \Omega_N^{2} \end{pmatrix}     - \omega^2\begin{pmatrix} 1+C_{11} & ... & C_{1N} \\ \vdots & \ddots & \vdots \\ C_{N1} & ... & 1+C_{NN} \end{pmatrix}  = 0
   \label{eq:HelmholtzPertubated_Syst}
\end{equation}
The eigensolutions of eq.~\eqref{eq:HelmholtzPertubated_Syst}, $(\tilde{\Omega}_i,|\tilde{\Psi}_i\rangle)_{i\in[1,N]}$, are the eigensolutions of eq.~\eqref{eq:Helmholtz} for the permittivity distribution $\tilde{\epsilon}(r)$.
In eq.~\eqref{eq:HelmholtzPertubated_Syst}, the coupling coefficients, $C_{ij}$, between original modes $i$ and $j$ depend on the variation of the permittivity and the spatial overlap of the modes at the location of the permittivity modification.
Noteworthily, the coupling integral not only depends on the spatial overlap of the mode intensity profiles but also on the overlap of their spatial distributions.
Remarkably, when reduced to two modes, the system is the analog of two inductance/capacitor oscillators coupled via an inductance $L_c$, in which charges of both capacitors satisfy
\begin{equation}
   \begin{pmatrix} \left(\frac{1}{\sqrt{L_1C_1}}\right)^2 & 0 \\  0 &  \left(\frac{1}{\sqrt{L_2C_2}}\right)^2 \end{pmatrix}   - \omega^2\begin{pmatrix} 1 + \frac{L_C}{L_1} &  \frac{L_c}{L_1} \\ \frac{L_c}{L_2} & 1 + \frac{L_c}{L_2}\end{pmatrix}  = 0 
   \label{eq:HelmholtzPertubated_SystLC}
\end{equation}
Eq.~\eqref{eq:HelmholtzPertubated_Syst} extends this result to any number of interacting modes $N>2$.
The modes act as a network of linearly coupled oscillators.

%% Figure 1 %%
\floatsetup[figure]{style=plain,subcapbesideposition=top}
\begin{figure}[h!]
   \includegraphics[scale=0.6]{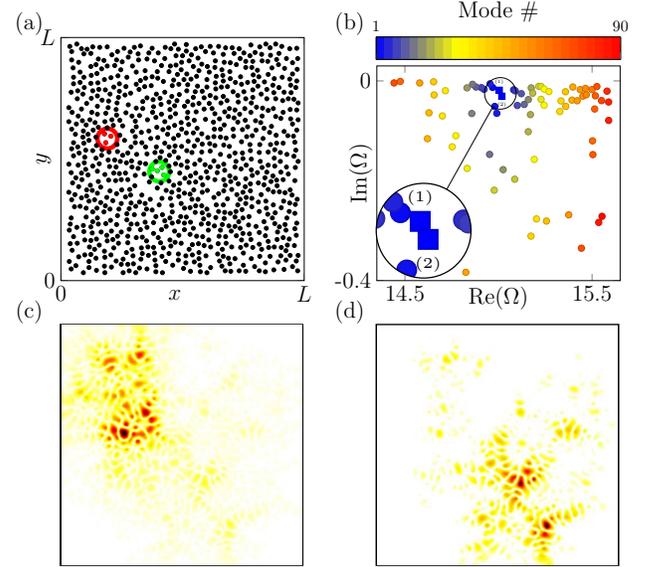}
    
\caption{
(a) 2D random medium: 896 scatterers of dielectric permittivity $\epsilon = 4$ are embedded in vacuum $\epsilon_{mat}=1.0$. 
The system is open at its boundaries. 
The permittivity is modified in two regions of diameter $340\, nm$ (respectively $R_1$ = red circle and $R_2$ = green circle). 
(b) Initial eigenvalues $(\Omega_i)_{i\in[1,90]}$ computed by FEM and sorted in the complex plane according the distance $d(1,i)$. An insert points out eigenvalues of interest (namely $\Omega_1$ and $\Omega_2$).
(c) and (d) Amplitudes of initial eigenvectors $|\Psi_1\rangle$ and $|\Psi_2\rangle$, respectively.
}
\end{figure}

Our theory is now applied to a particular system.
We consider a 2D random collection of 896 circular dielectric scatterers (radius 60 nm) with dielectric permittivity, $\epsilon = 4$, embedded in a host material of index $\epsilon_{mat} = 1.0$, with a filling fraction of $40\%$ (Fig.~1(a)).
The system dimensions are $L\times L = 5.3\, {\mu}m \times 5.3 \, {\mu}m$.
In the spectral range considered in the following, the localization length is estimated around $\xi \approx 1 \, {\mu}m \ll L$ and the modes are localized.
The two circular regions of diameter $340 \, nm$, $R_1$ and $R_2$, are shown in Fig.~1(a).
The dielectric permittivity of the scatterers within these regions is varied from $\epsilon$ to $\epsilon + \Delta \epsilon_1$ and $\epsilon + \Delta \epsilon_2$, respectively.
This can be achieved experimentally by shining 2 laser beams to induce a local change of the permittivity through nonlinear Kerr effect.

The initial modes $(\Omega_i,|\Psi_i\rangle)_{i\in[1,N]}$, which are the only input requested by eq.~\eqref{eq:HelmholtzPertubated_Syst} are computed using a Finite Element Method (FEM) \cite{RachidTouzani2013,Hernandez2005} with absorbing boundary conditions. 
Numerical boundary conditions are placed $0.4\, {\mu}m$ away from each side of the system.
A large number of modes ($N=90$) are computed for the initial system (Fig.~1(b)) in a narrow spectral range and we check that no modes are degenerated.
We choose among them two localized states $|\Psi_1\rangle$ and $|\Psi_2\rangle$ respectively at $\Omega_1$ and $\Omega_2$, spectrally close (Fig.~1(b)) but spatially distinct (Fig.~1(c),(d)).
We define in the complex plane the spectral distance of mode $i$ to mode $1$ as $d(1,i) = |\Omega_1 - \Omega_i|$.
This distance, color-coded in Fig.1(b) is a measure of the spectral overlap between mode $i$ and mode $1$.
Here, mode $2$ is most likely to couple to mode $1$ but we will see that the influence of other nearby modes cannot be neglected in the modal interaction.

The biorthogonal product defined in eq.~(\ref{eq:Projection}) corresponds to an integration over the whole space, $\mathbb{R}^2$
\begin{equation}
   \langle \Phi_q^* |\epsilon(r)| \Phi_p \rangle = \int_{\mathbb{R}^2} \epsilon(r) \Phi_q(r) \Phi_p(r) dr
   \label{eq:BiorthogonalProd}
\end{equation}
Since the initial modes are obtained via FEM computation, thus they are spatially defined in a finite spatial domain $V=[-0.1 \, {\mu}m,5.7\, {\mu}m]^2$.
However, the biorthogonal product can be split into an integral over $V$ and a second term at the boundary of $V$ which replaces outside propagation \cite{Doost2013a}. 
Because the modes are localized they have very weak amplitude along the boundary of $V$. 
Neglecting the edge term in the biorthogonal product leads to an inaccuracy of  0.8\% in the position of EP.
Therefore, we can approximate the biorthogonal product by its restriction to $V$.

%% Figure 2 %%
\floatsetup[figure]{style=plain,subcapbesideposition=top}
\begin{figure}[h!]
\sidesubfloat[]{\includegraphics[scale=.485]{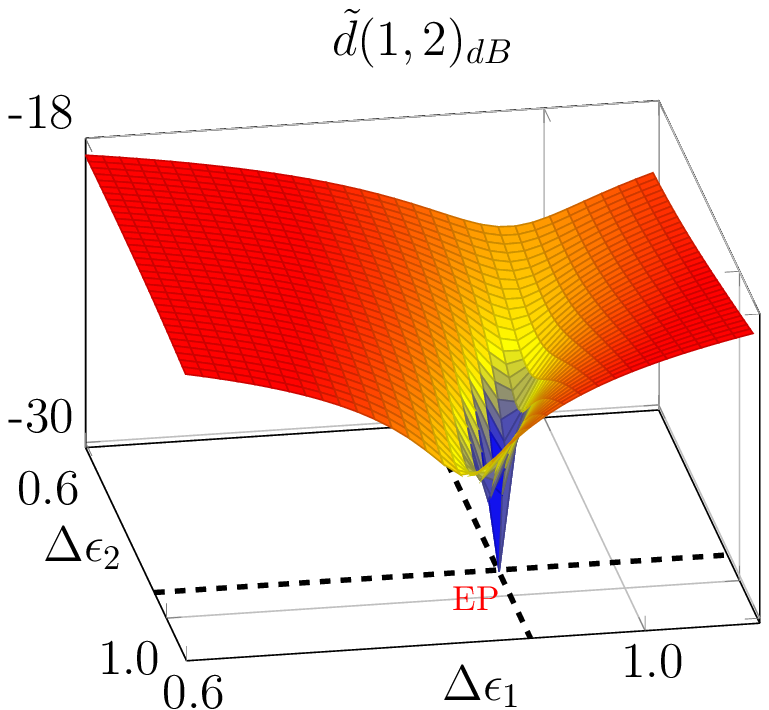}} \hfill
  \sidesubfloat[]{\includegraphics[scale=.485]{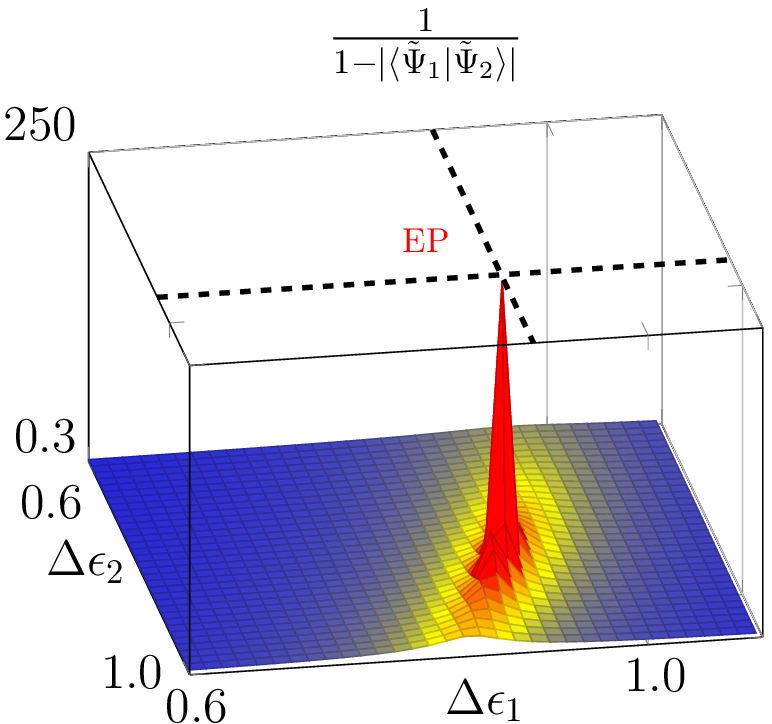}}
  
    \sidesubfloat[]{\includegraphics[scale=.485]{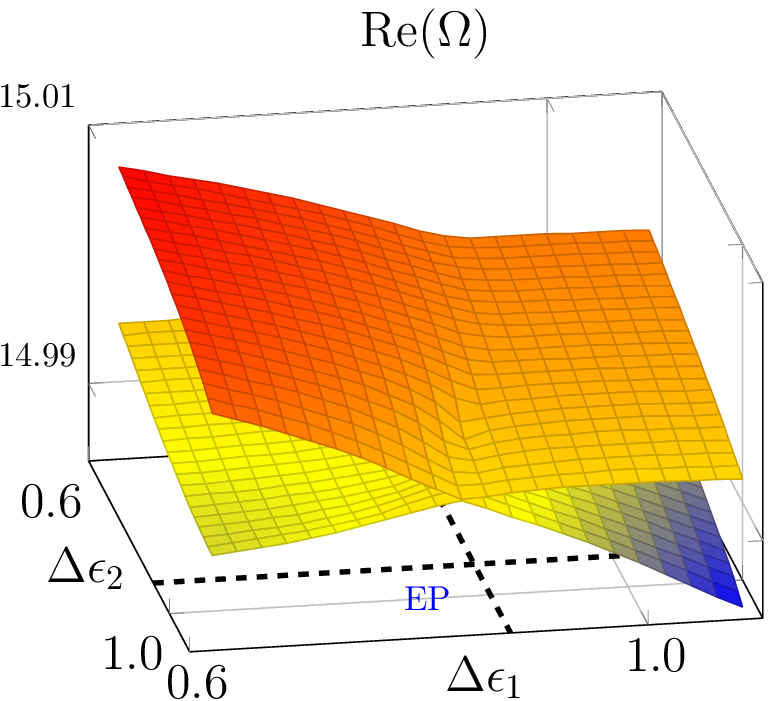}} \hfill
  \sidesubfloat[]{\includegraphics[scale=.485]{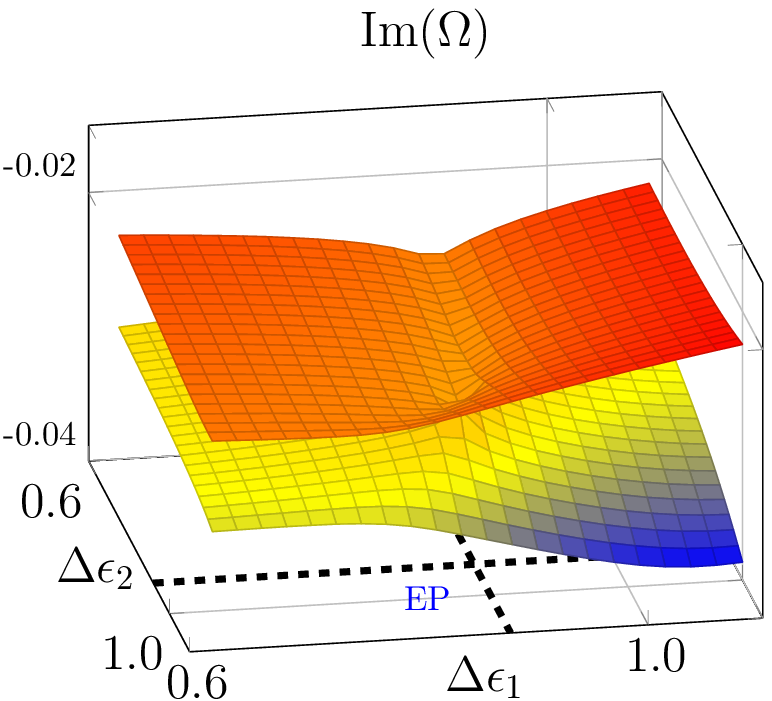}} \hfill
	 \sidesubfloat[]{\includegraphics[scale=1.0]{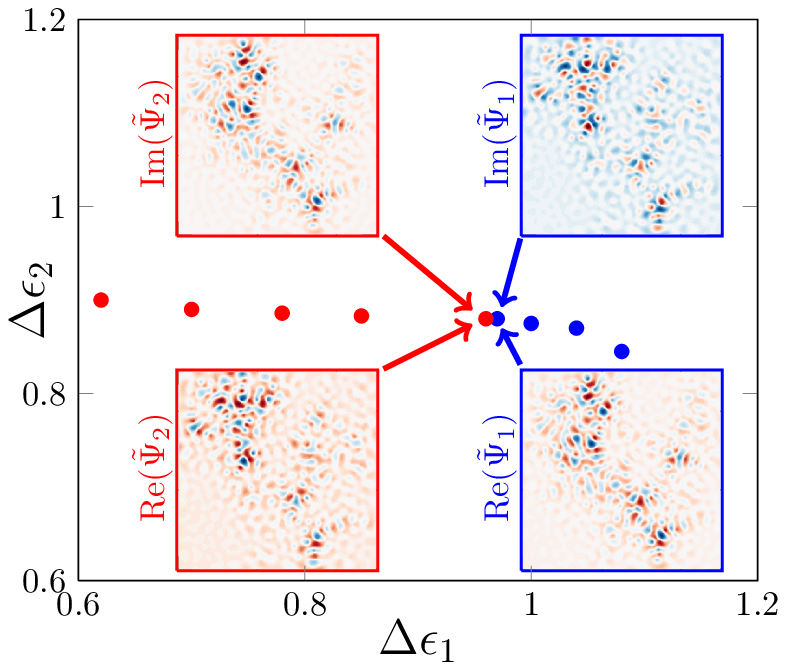}}

\caption{
(a) Eigenvalue difference $\tilde{d}(1,2)_{dB} = |\tilde{\Omega}_1 - \tilde{\Omega}_2|_{dB}$ (arbitrary units), canceling out at position $(\Delta \epsilon_1,\Delta \epsilon_2)_{@EP}$ in the parameter space. (b) The function $1/(1-|\langle \tilde{\Psi}_1|\tilde{\Psi}_2  \rangle|)$ in the parameter space. At position $(\Delta \epsilon_1,\Delta \epsilon_2)_{@EP}$, it exhibits a clear maximum meaning the two modes become collinear. (c) and (d) Real and imaginary parts of eigenvalues  $\tilde{\Omega}_{1}$ and $\tilde{\Omega}_{2}$. The eigenvalue surfaces display the typical structure of intersecting Riemann sheets centered around $(\Delta \epsilon_1,\Delta \epsilon_2)_{@EP}$. (e) Trajectories $\text{Im}(\tilde{\Omega}_1)= \text{Im}(\tilde{\Omega}_2)$ and $\text{Re}(\tilde{\Omega}_1)= \text{Re}(\tilde{\Omega}_2)$ marked by red and blue dots, respectively: Both curves join continuously at EP.
In inset, spatial distributions surrounded by a blue (red) frame show real and imaginary parts of mode 1 (2) at EP: Modes becomes collinear and satisfy $\tilde{\Psi}_1 = \pm i \tilde{\Psi}_2$.
}
\end{figure}

The parameter space, $\left(\Delta\epsilon_1,\Delta\epsilon_2\right)$, is sampled and eq.~(\ref{eq:HelmholtzPertubated_Syst}) is solved for each value of $\left(\Delta\epsilon_1,\Delta\epsilon_2\right)$ and for $N=60$ interacting modes to compute the new eigenstates, $(\tilde{\Omega}_i,|\tilde{\Psi}_i\rangle)_{i\in[1,N]}$.
The evolution of spectral distance between new modes 1 and 2, $\tilde{d}(1,2) =  |\tilde{\Omega}_1 - \tilde{\Omega}_2|$, in the small range of parameter space is shown in Fig.~2(a).
It sharply drops to zero at $(\Delta \epsilon_1,\Delta \epsilon_2)_{@EP} = (0.939,0.90)$.
The existence of an EP at this position is confirmed by plotting in the parameter space the real and imaginary parts of the eigenvalues, $\tilde{\Omega}_{i\in[1,2]}$ (Fig.~2(c),(d)). 
The intricate topology of intersecting Riemann's sheets around the singular point $(\Delta \epsilon_1,\Delta \epsilon_2)_{@EP}$, is the hallmark of an EP.
We now focus on the corresponding eigenfunctions, $\langle \tilde{\Psi}_1| \tilde{\Psi}_2 \rangle$, in the vicinity of the EP.
We compute the inner product of these two eigenstates and plot the function $1/(1-|\langle \tilde{\Psi}_1^{}| \tilde{\Psi}_2 \rangle|)$, as shown in Fig.~2(b).
At the exact position where $|\tilde{d}(1,2)|$ vanishes, the two eigenvectors become collinear confirming the coalescence of the two eigenstates.
The evolution of the modes in the vicinity of an EP is investigated in Fig.~2(e), where we plot the curves $\text{Im}(\tilde{\Omega}_1)= \text{Im}(\tilde{\Omega}_2)$ and $\text{Re}(\tilde{\Omega}_1)= \text{Re}(\tilde{\Omega}_2)$ in the parameter space.
As expected these two curves end at the exact position of the EP, where both the real and imaginary parts of the eigenvalues are equal, and they form a continuous trajectory.
Our calculations allows to go further and check that $|\tilde{\Psi}_{1} \rangle = \pm i |\tilde{\Psi}_{2} \rangle$ on either sides of the EP \cite{Gunther2007}.
Indeed, we confirm that both eigenvectors are collinear with a phase shift of $\pm \pi/2$ \cite{Gunther2007}.
Finally, the theoretical prediction is confronted with an numerical computation of the EP position: FEM simulations are used this time to compute systematically the new modes $|\tilde{\Psi}_{i\in [1,2]} \rangle$ for each set of $(\Delta \epsilon_1,\Delta \epsilon_2)$ in a range  enclosing our theoretical prediction $(\Delta \epsilon_1,\Delta \epsilon_2)_{@EP}$.
The sampling is set to $0.04$ along $\Delta \epsilon_1$ and $\Delta \epsilon_2$.
The numerically-computed eigenvalues merge at position $(\Delta \epsilon_1,\Delta \epsilon_2) = (0.92,0.88) \pm (0.04,0.04)$ which is within the error bar of the predicted value obtained from eq.~\eqref{eq:HelmholtzPertubated_Syst}. 
The collinearity of the eigenvectors is also satisfied and confirms the existence of an EP at this specific position.

In order to understand the coalescence mechanism, the common amplitude of mode 1 and 2 at EP, $|\tilde{\Psi}_{1/2,@EP}\rangle$, is shown in the inset of Fig.~3(a).
It forms a beaded chain which connects both ends of the system and is similar to necklace states studied in \cite{Labonte2012}.
To investigate the influence of the nearby modes in the coalescing of modes 1 and 2 at EP, we measure in Fig.~3(a) the norm of the biorthogonal projection of eigenstate $|\tilde{\Psi}_{1/2,@EP}\rangle$ along the different initial modes, $| \langle \Psi_{i\in [1,N]} | \tilde{\Psi}_{1/2,@EP} \rangle |$.   
Remarkably enough, it demonstrates the vanishing, though non negligible, influence of nearby modes, $i>2$.
If projection norms for initial modes 1 and 2 are close to 0.6, others modes (for instance 4 and 5) contribute up to 0.25.    
Their influence is also highlighted in Fig.~3(b), where the position of the EP, $(\Delta \epsilon_1,\Delta \epsilon_2)_{@EP}$, is obtained from eq.~\eqref{eq:HelmholtzPertubated_Syst} for different number of modes, $N$, ranging from 2 to 60.
The position of the EP in the parameter space fluctuates significantly for small $N$ and converges for values larger than 55.
Therefore, the EP cannot be explained by a two-mode interaction and results from a collective effect between multiple modes.

%% Figure 3 %%
\begin{figure}[h!]

   \sidesubfloat[]{\includegraphics[scale=.9]{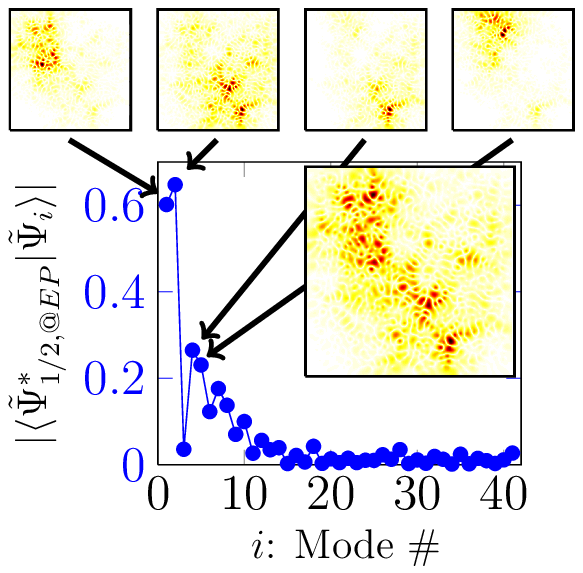}} 
   
   \sidesubfloat[]{\includegraphics[scale=.7]{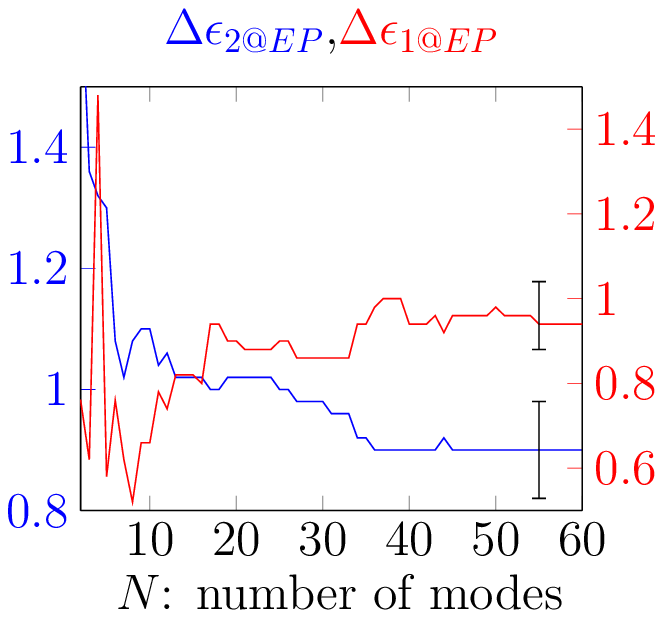}}

\caption{
(a) Norm of the biorthogonal projection of modes 1 and 2 at EP along the 45 first initial modes: $| \langle \Psi_{i\in [1,N]} | \tilde{\Psi}_{1/2,@EP} \rangle |$. The amplitude of $|\tilde{\Psi}_{1/2,@EP}\rangle$ is shown in inset. 
The four aligned spatial distributions correspond to initial modes 1, 2, 4 and 5, respectively. Black arrows point to their respective biorthogonal projection coefficients. 
(b) Prediction of EP position in the parameter space ($\Delta \epsilon_{1@EP}$ red curve and $\Delta \epsilon_{2@EP}$ blue curve) for a number of initial modes $N$ ranging from 2 to 60. Prediction only converges for $N \ge 55$. The two error bars point out the position of the EP obtained by FEM computation.
}
\end{figure}

In conclusion, we proposed a general theory to study the evolution of modes in open media when the permittivity is varied.
This approach, which relies on the linearity in $\epsilon(r)$ in Helmhotz equation, is non-perturbative: For any $\Delta \epsilon_{i\in [1,2]}$,  eq.~\eqref{eq:HelmholtzPertubated_Syst} gives the new modes from the knowledge of the initial system.
Our theory describes the system by an infinite set of modes acting as  oscillators coupled via the modification of the permittivity.
We considered the specific case of random media in the localization regime
and show that our theory can be used to investigate the mode coupling and hybridization resulting from a local perturbation.
In particular, by changing the local index at two different locations, two modes are brought to an EP where they coalesce.
Remarkably, the accuracy of the theoretical prediction is shown to strongly depend on the number of modes considered.
A large number of modes is required, which shows that the evolution of modes is dictated by multiple-mode interactions.
In term of experimental realization, such a disorder manipulation can be easily implemented on existing setups \cite{Riboli2011,Yang2011,Riboli2014}.
Permittivity landscape can be shaped reversibly using e.g. laser illumination and  nonlinear Kerr effect to engineer the modes.
For instance, EP can be reached by following the crossing line of Fig.~2(e). 
EPs can be calculated for any pair of modes and even generalized to three or more eigenstates coupling, which opens the way to the control of light-matter interaction in random media.
For instance, the random system can be designed to force the hybridization of modes and the opening of channels in opaque systems. 
This is in contrast to earlier works where such a necklace state was obtained using a try-and-fail method.
Alternatively, using mode repulsion in the vicinity of EP, the disorder can be engineered to increase the spatial confinement of the modes and consequently their Q-factor.
Finally, we believe this approach can be extended to others types of cavity networks e.g. photonic crystal cavity arrays used for quantum simulation \cite{Hartmann2006,Greentree2006,Majumdar2012}, where fabrication inaccuracy could be compensated \textit{a posteriori} by such an external control. 

We acknowledge S. Rotter for fruitfull discussions. PS is thankful
to the Agence Nationale de la Recherche for its support under grant ANR PLATON (No. 12-BS09-003-01), the LABEX WIFI (Laboratory of Excellence within the French Program `Investments for the Future') under reference ANR-10-IDEX-0001-02 PSL* and the Groupement de Recherche 3219 MesoImage.

%
%\bibliography{Article_PE}

\end{document}